\documentclass{article}
\usepackage {graphicx}
\begin{document}
\baselineskip .75cm 
\begin{titlepage}
\title{\bf Generalized Cauchy-Schwarz inequality and Uncertainty Relation}       
\author{Vishnu M. Bannur  \\
{\it Department of Physics}, \\  
{\it University of Calicut, Kerala-673 635, India.} }   
\maketitle
\begin{abstract}
  A generalized Cauchy-Schwarz inequality is derived and applied to uncertainty relation in quantum mechanics. We see a modification in the  uncertainty relation and minimum uncertainty wave packet. 
\end{abstract}
\vspace{1cm}
                                                                                
\noindent
{\bf PACS Nos :}  03.65.Ta 03.67.-a 03.65.Ca \\
{\bf Keywords :} Cauchy-Schwarz inequality, uncertainty relations, minimum uncertainty wave packets.
\end{titlepage}
\section{Introduction :}

The uncertainty relation is an inherent property of quantum mechanics which is visible in atomic and subatomic world. The origin of this property was debated for so many years in the literature without any clear conclusions and lead to different interpretations of quantum mechanics, theories like Hidden variables, etc. \cite{th.1}. One of the standard method to derive the  uncertainty relation is based on Cauchy-Schwarz inequality. 

Here, we revisit Cauchy-Schwarz inequality (CS) and generalize it. Then we reanalyze the derivation of uncertainty relation and it's consequences. Recently, there has been renewed interest in uncertainty relation \cite{hu.1, pa.1, ba.1, so.1, li.1} because of it's relevance to entanglement, cryptography, signal processing, squeezed states, etc. 

\section{Cauchy-Schwarz inequality and Uncertainty relations:} 

In our ordinary vector analysis, we know that,
\begin{equation}
  \vec{A}.\vec{B} = A B \cos \theta \,\,\,,
\end{equation} 
 and hence, 
\begin{equation}
   |\cos \theta | = \frac{| \vec{A}.\vec{B}|}{A B} \le 1, \,\,\, \mbox{or} \,\,\, A B \ge |\vec{A}.\vec{B}| \,\,\, ,
\end{equation}
 which is known as Cauchy-Schwarz inequality. Generalizing it to higher dimension such as such as quantum mechanics Hilbert space, we may write $\|A\|\, \|B\| \ge |<A|B>|$ where $\|A\|$ is the norm of vector $|A>$ defined as $\| A\|  \equiv \sqrt{<A|A>} $ and $<A|B>$ is the scalar product of two states $|A>$ and $|B>$. Following the standard method, we define a vector $|X> = |A> + \lambda |B>$ and it is clear that $\|X\|^2 \ge 0$. That is,
\begin{equation}
  <X|X> = (<A| + \lambda^* <B|) (|A> + \lambda |B>) \ge 0 \,\,\,,
\end{equation}
or,
\begin{equation}
  \|A\|^2 + \lambda \lambda^* \|B\|^2  + \lambda <A|B> + \lambda^* <B|A> \ge 0 \,\,\,, \label{eq:cs1} 
\end{equation}
where $\lambda$ is a free parameter and we may fix it by minimizing the left hand side of above equation and get,  
\begin{equation}
   \lambda^* \|B\|^2  + <A|B>  =  0 \,\,\,,  
\end{equation}
and hence 
\begin{equation}
   \lambda^* = -  \frac{<A|B>}{\|B\|^2}  \,\,\,.  
\end{equation}
With this $\lambda$, inequality equation, Eq.(\ref{eq:cs1}),  reduces to 
\begin{equation}
  \|A\|^2 \|B\|^2 \ge |<A|B>|^2 \,\,\,, \label{eq:cs2} 
\end{equation}
the CS inequality. 

In quantum mechanics with canonical quantization rule, we have Hermitian operators and vectors in Hilbert space \cite{th.1, me.1}. The observable corresponding to any Hermitian operator may be obtained by taking the expectation values. Expectation values like average values which can not be obtained by single measurement or single system. Different measurements will give different results and deviates from mean value because of the intrinsic property of quantum mechanics. Hence, we define the deviation $\Delta A \equiv <\psi| (A - \bar{A})|\psi> = 0$, where $\bar{A} \equiv <A> = <\psi|A|\psi>$ and $|\psi>$ is the state of the system. So we take mean square deviation $<\psi| (A - \bar{A})^2|\psi>  = <A^2> - <A>^2 \equiv \Delta A^2$, square root of which may be taken as the measure of deviation from the mean value, or called uncertainty in the observable $A$. 

According to uncertainty relation, there exists always a relation between the uncertainties of observables corresponding to incompatible operators, first pointed out by Heisenberg \cite{he.1}. For example, uncertainty in position and momentum, among the components of angular momentum, etc. It follows from CS inequality as given below.  

Let $A$ and $B$ are two incompatible operators such that commutator $[A,B] = i C$. By choosing $|\psi_A>  \equiv (A - \bar{A}) |\psi>$ and $|\psi_B>  \equiv (B - \bar{B}) |\psi>$, CS inequality leads to, 
\begin{equation}
 \|\psi_A\|^2 \|\psi_B\|^2 \ge |<\psi_A|\psi_B>|^2 \,\,\,.
\end{equation} 
But $\|\psi_A\|^2 = <\psi| (A - \bar{A}) (A - \bar{A}) |\psi> = \Delta A^2 $ and $\|\psi_B\|^2 = \Delta B^2$, and hence,
\begin{equation}
 \Delta A^2 \Delta B^2 \ge |<\psi_A|\psi_B>|^2 \,\,\,, \label{eq:sh}  
\end{equation} 
which may be simplified to get,  
\begin{equation}
  \Delta A ^2 \Delta B ^2 \ge  | Im(<\psi_A |\psi_B>)|^2  = \frac{1}{4}|<[A,B]>|^2\,\,\, , \label{eq:hr}
\end{equation}
where $Im$ refers to imaginary part. This is known as Heisenberg-Robertson (HR) uncertainty relation \cite{ro.1} which leads to familiar uncertainty relation for position and momentum $\Delta x$ $\Delta p$ $\ge \frac{1}{2}|<[x,p]>| = \frac{\hbar  }{2} $. If we keep both real and imaginary parts of the left hand side (LHS) of Eq. (\ref{eq:sh}), we get Heisenberg-Robertson and Schrodinger (HRS) \cite{sc.1} uncertainty relation, stronger than HR uncertainty relation,  
\begin{equation}
  \Delta A ^2 \Delta B ^2 \ge  \frac{1}{4}|<[A,B]>|^2 + \frac{1}{4} |<\{A,B\}> - 2 <A> <B>|^2,
\end{equation}
where $\{A,B\}$ is anti-commutator. 

\section{Generalized CS inequality:}  

In the earlier section we used the fact that  $\|X\|^2 \ge 0$ and equality is for $|X> = 0$, a null vector. In general, $|X> = \sum_n C_n |n>$ where $|n>$ is a orthogonal basis. Therefore, $\|X\|^2 = \sum_n |C_n|^2 = |C_1|^2 + |C_2|^2 + ...$ and hence the inequality may be made stronger by noticing that $\|X\|^2 \ge  |C_m|^2$ where $C_m$ may be taken as the probability amplitude which has largest probability. Only when all $C_n$s are zero, we have $|X> =0$ or $\|X\|^2 \ge 0$ and $\|X\|^2 \ge |C_m|^2$ is a more general and stronger inequality. 

Substituting for $|X> = |A> + \lambda |B>$ and $C_m = <m|A> + \lambda <m|B> \equiv a_m + \lambda b_m$, we get, 
\begin{equation}
  (\|A\|^2 - |a_m|^2) + \lambda \lambda^* (\|B\|^2 - |b_m|^2)  + \lambda (<A|B> - b_m a_m^*) + \lambda^* (<B|A> - b_m^* a_m) \ge 0 \,\,\,, \label{eq:cs3} 
\end{equation}
Again the minimum value is 
\begin{equation}
   \lambda^* = -  \frac{(<A|B> - b_m a_m^*)}{(\|B\|^2 - |b_m|^2)}  \,\,\,.  
\end{equation}
and a new inequality is 
\begin{equation}
  (\|A\|^2 - |a_m|^2)  (\|B\|^2 - |b_m|^2) \ge |(<A|B> - b_m a_m^*)|^2 \,\,\,, \label{eq:cs4} 
\end{equation}
Clearly for $|X> = 0$ or for $C_m = 0$, it reduces to the standard CS inequality. 

Proceeding the same way as earlier section, a new uncertainty relation may be obtained by choosing $|\psi_A>  \equiv (A - \bar{A}) |\psi>$ and $|\psi_B>  \equiv (B - \bar{B}) |\psi>$, etc. for incompatible operators $A$ and $B$, as 
\begin{equation}
  (\Delta A^2 - |a_m|^2) (\Delta B^2 - |b_m|^2) \ge |(<\psi_A|\psi_B> - b_m a_m^*)|^2
\end{equation}
It should be pointed out that the inequality equation, Eq.(\ref{eq:cs3}), before minimization, was studied recently by Maccone and Pati \cite{pa.1}  for $\lambda = \pm i, \pm 1$ and claimed to be stronger uncertainty relations. However, we point out that those are just inequalities only and not the uncertainty relations \cite{ba.1}. Further, those inequalities are meaningless for incompatible operators with different dimensions like position and momentum. Such inequalities are dimensionally wrong. 
 
\section{Minimum uncertainty wave packets: } 

It is well known that the wave packet or wave function corresponding to minimum uncertainty is Gaussian for a linear harmonic oscillator (HO). It is based on the fact that $|X> = 0$, where $|X> = |\psi_A> + \lambda |\psi_B> = 0$ Here, $A = x$ and $B=p$ and $\bar{x} = \bar{p} = 0$ for HO. Hence, $x |\psi> + \lambda p |\psi> = 0$, which on coordinate representation is,
\begin{equation}
  x \psi(x) + \lambda (-i \hbar   ) \frac{d \psi(x)}{dx} = 0 \,\,\,,
\end{equation}
which immediately gives,
\begin{equation}
  \psi(x) = C \exp (\frac{-x^2}{2 a^2}) \,\,\, ,
\end{equation}
where $a^2 \equiv -i \hbar \lambda $. 
For minimum uncertainty, $\lambda$ is given by, 
\begin{equation}
   \lambda = - \frac{<B|A>}{\Delta B^2} = - \frac{<\psi|p x|\psi>}{\Delta p^2} \,\,\,,  
\end{equation}
which on evaluation in coordinate space, we get, 
\begin{equation}
  \lambda = \frac{i \hbar }{\Delta p^2} (1 + \epsilon) \,\,\,, 
\end{equation}
where
\begin{equation}
  \epsilon = \int dx \psi^* x \frac{d \psi}{dx} = - \frac{1}{2} \,\,\,,
\end{equation}
and 
\begin{equation}
  \lambda = \frac{i \hbar }{2 \Delta p^2} \,\,\,, 
\end{equation}
and hence $ a^2 = \frac{\hbar ^2}{2 \Delta p^2}$, and 
\begin{equation}
  \psi(x) = C \exp (\frac{-x^2}{4 \Delta x^2}) \,\,\, ,
\end{equation}
where we made use of minimum uncertainty relation $\Delta x^2 \Delta p^2 = \frac{\hbar^2 }{4}$. 
After normalizing the wave function, we get, 
\begin{equation}
  \psi(x) = \frac{1}{(2 \Delta x^2 \pi )^{1/4}} \exp (\frac{-x^2}{4 \Delta x^2}) \,\,\, , 
\end{equation}
a well known expression in the standard text book. 

Applying the same procedure for our case of general uncertainty relation, we have, $|X> = x_m |m> $ for the minimum uncertainty, or 
\begin{equation}
  x |\psi> + \lambda p |\psi> = x_m |m> = (a_m + \lambda b_m) |m> \,\,\,,
\end{equation}
which may be solved to get,
\begin{equation}
  \psi(x) = C \exp (\frac{-x^2}{2 a^2}) + (a_2 + \frac{a_1}{a^2}) \exp (\frac{-x^2}{2 a^2}) f(x) \,\,\, , \label{eq:wf} 
\end{equation}
where 
\begin{equation}
  f(x) \equiv \int^x dy u_m(y) 
  \exp (\frac{y^2}{2 a^2})\,\,\, ,
\end{equation}
\begin{equation}
  a_1 \equiv a_m = \int dx x u_m(x) \psi(x) \,\,\, , 
\end{equation}
\begin{equation}
  a_2 \equiv \frac{b_m}{-i \hbar } = \int dx u_m(x) \frac{d}{dx} \psi(x)\,\,\, ,
\end{equation}
and $u_m(x) = <x|m>$. Since $|X>$ is orthogonal to $| \psi>$, i.e., $<\psi|X> = 0$, all $|n>$ states in $|X> = \sum_n x_n |n>$ are orthogonal to $|\psi>$. Therefore, as an example, let us choose $u_m (x) = N x \exp (-\alpha x^2)$ which on normalization gives,
\begin{equation}
  u_m(x) = \left( \frac{32 \alpha^3}{\pi}\right) ^{1/4} x \exp (- \alpha x^2) \,\,\,. 
\end{equation}
Then we may evaluate f(x), $a_1$, $a_2$, etc., and finally we get,
\begin{equation}
  \psi(x) = C \exp (\frac{-x^2}{2 a^2}) + \left[ a_1 \left( \frac{32 \alpha^3}{\pi}\right) ^{1/4} - C \sqrt{\frac{8}{(1+\frac{1}{2 a^2 \alpha} )^3}}\right] \exp (- \alpha x^2)\,\,\, ,
\end{equation}
where $C$ may be fixed by normalization of $\psi(x)$ and $a_1$ still unknown. First term is the standard minimum uncertainty relation part and second term is the new modification which happened to be Gaussian. The width of the first Gaussian is determined by $a^2 \equiv -i \hbar \lambda$ which may be evaluated as,  
\begin{equation}
  a^2 = \frac{\Delta^2_A}{\bar{a}^2} \,\,\,,
\end{equation}
where $\Delta^2_A \equiv (\Delta A^2 - |a_1|^2)$ and $\bar{a}^2 \equiv (\frac{1}{2} - a_1 a_2)$. Two equations for $a_1$ and $a_2$ reduce to the same equation on simplification so that we are unable to solve them for $a_1$ and $a_2$, but are related. Interesting point is that we see a modification of the width of the standard minimum uncertainty wave packet which may be squeezed or stretched depending on the values of $a_1$, $C$ and $\alpha$.    

\section{Conclusions:} 

 Cauchy-Schwarz inequality and uncertainty relation in quantum mechanics is revisited, and generalized to a stronger inequalities. As an example, one consequence of the generalization is that the minimum uncertainty wave packet is modified with an additional function and the standard Gaussian minimum wave packet squeezes or stretches depending on the parameters of the system and measurements.

\end{document}